\documentclass[aps,prb,reprint,nofootinbib,superscriptaddress]{revtex4-2}

\usepackage[utf8]{inputenc}
\usepackage[T1]{fontenc}
\usepackage{lmodern}
\usepackage{amsmath,amssymb,bm}
\usepackage{graphicx}
\usepackage{dcolumn}
\usepackage{xcolor}
\usepackage[normalem]{ulem}
\usepackage[colorlinks=true,linkcolor=blue,citecolor=blue,urlcolor=blue]{hyperref}
\hypersetup{pdftitle={High-density correlation energy of the electron gas},
  pdfauthor={Miguel A. L. Marques, Paulo R. Pires, Stefano Pittalis}}
\usepackage{orcidlink}

\DeclareRobustCommand{\SP}[2]{%
  \textcolor{blue}{%
    \if\relax\detokenize{#1}\relax\else
      \sout{#1}%
      \if\relax\detokenize{#2}\relax\else\space\fi
    \fi
    #2}}

\newcommand{\valA}{0.031091}

\newcommand{\valCRPA}{0.002710}
\newcommand{\valCX}{0.006520}
\newcommand{\valBx}{0.024179}
\newcommand{\EtwoT}{-0.0868}
\newcommand{\EtwoIn}{5.069}
\newcommand{\EtwoAL}{-6.415}
\newcommand{\Etwo}{-1.433}
\newcommand{\EtwoLong}{-1.432664}
\newcommand{\DRPA}{-0.0072147}
\newcommand{\DRlnA}{-0.0011118}
\newcommand{\DRlnh}{-0.0009885}
\newcommand{\DRhard}{-0.0051145}
\newcommand{\DXone}{-0.0084606}
\newcommand{\DXlnA}{-0.0026752}
\newcommand{\DXlnh}{-0.0021265}
\newcommand{\DXhard}{-0.0036589}
\newcommand{\Idir}{970.47(10)}
\newcommand{\Iex}{4.86(4)}

\newcommand{\Ethree}{0.0078874(9)}
\newcommand{\Dnew}{-0.0077880(9)}

\newcommand{\Dold}{-0.010}
\newcommand{\Eendo}{0.0075}
\newcommand{\cmepspp}{-0.0105}
\newcommand{\DRPAendo}{-0.0072}
\newcommand{\cmlnA}{-0.00485}
\newcommand{\cmlnh}{-0.00185}
\newcommand{\cmhard}{-0.004}
\newcommand{\Atwelvethird}{0.0021732}

\newcommand{\cmlnhLow}{13}
\newcommand{\cmNumDiff}{7\times10^{-5}}
\newcommand{\EthreeShift}{4\times10^{-4}}
\newcommand{\DchangeFrac}{one fifth}
\newcommand{\Eex}{1.1926}
\newcommand{\ETU}{0.3463}
\newcommand{\ECt}{0.3166}
\newcommand{\Ein}{-0.6581}
\newcommand{\EAL}{1.7754}
\newcommand{\Eit}{0.8111}
\newcommand{\Exx}{-0.1307}
\newcommand{\Eone}{3.653(10)}
\newcommand{\EoneLong}{3.653}
\newcommand{\EoneNum}{0.0007}
\newcommand{\EoneMeth}{0.010}
\newcommand{\EoneRPA}{1.4849}

\newcommand{\cALzero}{-0.5366}
\newcommand{\bzero}{0.2858(3)}

\newcommand{\dmcAboveD}{2.2}

\newcommand{\dmcSplit}{1.1}
\newcommand{\pwHigh}{0.3}

\newcommand{\rs}{r_s}
\newcommand{\qtf}{q_\mathrm{TF}}
\newcommand{\atf}{a_\mathrm{TF}}
\newcommand{\Ctot}{C_\mathrm{tot}}
\newcommand{\CAL}{C_\mathrm{AL}}
\newcommand{\cAL}{c_\mathrm{AL}}
\newcommand{\bin}{b_\mathrm{in}}
\newcommand{\Ha}{\,\mathrm{Ha}}
\newcommand{\avg}[1]{\langle #1\rangle}
\newcommand{\dd}{\mathrm{d}}
\newcommand{\bochum}{Research Center Future Energy Materials and Systems and ICAMS, Ruhr University Bochum, Germany}

\begin{document}

\title{High-density correlation energy of the electron gas}

\author{Miguel A. L. Marques\orcidlink{0000-0003-0170-8222}}
\email{miguel.marques@rub.de}
\affiliation{\bochum}
\author{Paulo R. Pires\orcidlink{0009-0005-7598-4862}}
\affiliation{\bochum}
\author{Stefano Pittalis\orcidlink{0000-0001-6941-5579}}
\affiliation{Istituto Nanoscienze -- CNR, S3, Via G. Campi 213/A, 41125 Modena, Italy}
\date{\today}

\begin{abstract}

The high-density expansion of the correlation energy per electron of the unpolarized three-dimensional electron gas,
$\epsilon_c=A\ln\rs+B+C\rs\ln\rs+D\rs+\dots$, is used in many approximations to the correlation functional of
density-functional theory. The coefficients $A$, $B$ and $C$ are known exactly. We revisit the three diagram classes
used to estimate $D$, namely bare rings, rings with one first-order exchange insertion, and third-order diagrams with three
distinct momentum transfers. Their sum is $\Dnew$\,Ha per electron, compared with the earlier estimate of
$\Dold$\,Ha. The difference comes mainly from the coupling-constant weighting of the exchange rings in the
calculation of Carr and Maradudin. At order $\rs^2$, seven contributions to the single-logarithm coefficient $E_1$
sum to $(\EoneLong\pm\EoneMeth)\times10^{-3}$\,Ha per electron.
The three long-wavelength contributions to the double-logarithm coefficient $E_2$ sum to
$\alpha^2(36\pi^2-208\ln2-221)/(192\pi^4)\approx\Etwo\times10^{-4}$\,Ha per electron, where
$\alpha=(4/9\pi)^{1/3}$. Available diffusion Monte Carlo energies at $\rs=0.5$ cannot distinguish between the two
estimates of $D$.

\end{abstract}

\maketitle

\section{Introduction}

The uniform electron gas is the reference system of density-functional theory. Most approximations to the correlation
functional use its correlation energy per electron, $\epsilon_c(\rs)$, as a function of the Wigner--Seitz radius
$\rs$~\cite{Marques2012,Lehtola2018,Tran2026}. The parametrizations in use interpolate between quantum Monte Carlo
energies at intermediate densities and the analytic behavior at high density~\cite{Vosko1980,Perdew1981,Perdew1992,
Chachiyo2016,Xie2021}. In the high-density limit, $\rs\to0$, the correlation energy has the expansion
\begin{multline}
  \epsilon_c = A\ln\rs + B + C\rs\ln\rs + D\rs\\
  + E_1\rs^2\ln\rs + E_2\rs^2\ln^2\rs + O(\rs^2).
  \label{eq:expansion}
\end{multline}
Gell-Mann and Brueckner summed the ring diagrams in 1957 and obtained $A$ and the ring part of $B$~\cite{GellMann1957}.
The exchange part of $B$ was given in closed form by Onsager, Mittag and Stephen~\cite{Onsager1966}, and the ring part
by Hoffman~\cite{Hoffman1992}. Carr and Maradudin computed the terms of order $\rs$ in
1964~\cite{Carr1964}. Their coefficient $C$ was later confirmed and written in closed form by Loos and
Gill~\cite{Loos2011}. Determining $D$ also requires the energy $E_{32}$ of the third-order diagrams in which all three Coulomb lines
have different momenta. Endo \emph{et al.}\ evaluated these in 1999 and obtained $D\approx\Dold$\,Ha~\cite{Endo1999}.

However, the status of $D$ is unsatisfactory. Its largest part comes from the series of ring diagrams that contain
one first-order exchange insertion, and this part is taken from the numerical integrations of Carr and Maradudin.
Loos and Gill considered the value to be probably not accurate~\cite{Loos2016}. Beyond order $\rs$ nothing is known
in three dimensions. The expansion~\eqref{eq:expansion} through $\rs\ln\rs$ has one logarithm per order, and it is
tempting to assume that this pattern continues. Imoto recently organized the expansion with Mellin transforms, which
makes it possible to read off where logarithms arise, and listed the terms of order $\rs^2$ in three dimensions as an
open problem~\cite{Imoto2026}. Our group has studied the high-density limit of the gas in arbitrary
dimensions~\cite{Schlesier2020,Pham2023}.

In this work we evaluate contributions to $D$ and to the logarithmic terms of order $\rs^2$. We start from the ring sum over proper
polarization insertions and separate each momentum integral into a soft region, $q\sim\qtf$, and a hard region,
$q\sim k_F$, where $q$ is the momentum transfer, $\qtf$ the Thomas--Fermi wave vector, and $k_F$ the Fermi wave vector. We evaluate the three classes of diagrams used in the earlier estimate of $D$ and identify the source of the difference. The evaluated logarithmic contributions at order $\rs^2$ follow from the long-wavelength structure of the
first- and second-order polarization at fixed ratio of frequency to momentum. Finally, we compare the expansion with
diffusion Monte Carlo (DMC) energies.

\section{The ring sum and its regions}
\label{sec:ring}

\subsection{Units and the ring formula}

Throughout this work, we use Hartree atomic units, with $\rs$ in bohr, and $k_F=m=\hbar=1$ inside the diagrams, with momenta measured in units of $k_F$. The dimensionless coupling
is $e^2=\alpha\rs$ with $\alpha=(4/9\pi)^{1/3}$, the bare interaction is $v(q)=4\pi e^2/q^2$, and the Thomas--Fermi wave
vector is
\begin{equation}
  \qtf^2=\frac{4e^2}{\pi}=\atf\rs,\qquad \atf=\frac{4\alpha}{\pi}.
\end{equation}
All polarizations are taken at imaginary frequency $i\nu$, and we write $\nu=qz$. The ratio $z$ is the velocity of the
density fluctuation in units of the Fermi velocity. The free polarization of both spins is
$\Pi^{(0)}(q,iqz)=-(\qtf^2/4\pi e^2)L(q,z)$, where the Lindhard function has the expansion
\begin{equation}
  L(q,z)=h(z)+q^2L_1(z)+q^4L_2(z)+\dots
  \label{eq:lindhard}
\end{equation}
at fixed $z$. Its long-wavelength limit is $h(z)=1-z\arctan(1/z)$, the first correction is $L_1=-1/[12(1+z^2)^2]$,
and all $L_k$ are given in Appendix~\ref{app:lindhard}. The expansion contains no $\ln q$.

\begin{figure*}
  \includegraphics[width=\textwidth]{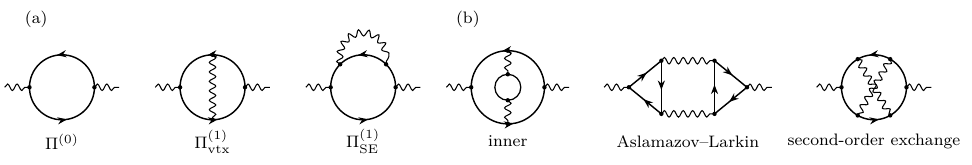}
  \caption{Proper polarization insertions of the ring sum, Eq.~\eqref{eq:ring}. Solid lines are free electron
  propagators and wavy lines Coulomb interactions. (a) The free polarization $\Pi^{(0)}$ and the two parts of the
  first-order exchange polarization, the vertex correction and the Fock self-energy. (b) Second-order polarization insertions. In the inner class a bubble screens the internal line of $\Pi^{(1)}$. The Aslamazov--Larkin
  diagrams connect two triangles by two interaction lines. The genuine second-order exchange diagrams have no
  logarithm.}
  \label{fig:diagrams}
\end{figure*}

Let $\Pi=\Pi^{(0)}+\Pi^{(1)}+\Pi^{(2)}+\dots$ be the proper polarization ordered in powers of $e^2$
[Fig.~\ref{fig:diagrams}]. The screening function is
\begin{equation}
  \mathcal C(q,z)=-v(q)\Pi(q,iqz)=c+c_1+c_2+\dots,
\end{equation}
with $c=(\qtf^2/q^2)L(q,z)$ from the free polarization. The energy of all diagrams that can be cut into a ring of proper
insertions is
\begin{equation}
  \epsilon=\frac{3E_0}{8\pi}\int_{-\infty}^{\infty}\!\dd z\int_0^\infty\!\dd q\,q^3\Big[\ln(1+\mathcal C)-\mathcal C\Big],
  \label{eq:ring}
\end{equation}
with $E_0=16/(\pi^2\qtf^4)$\,Ha. Every $z$ integral in this paper runs over the full real axis, and all integrands are
even in $z$. The subtraction of $\mathcal C$ removes the terms linear in the insertions, which are not ring diagrams.
These are the first-order exchange energy, the second-order exchange energy, and the closing of $\Pi^{(2)}$ with a single
line. The second-order exchange energy, $B_x=\ln2/6-3\zeta(3)/(4\pi^2)\approx\valBx$\,Ha~\cite{Onsager1966}, is part
of $B$. The closing of $\Pi^{(2)}$ belongs to the third-order energy $E_{32}$ of Sec.~\ref{sec:D}.

With $\mathcal C=c=h\qtf^2/q^2$ and $q=\qtf t$, the integrand of Eq.~\eqref{eq:ring} decays like $-h^2/(2t)$ at large $t$.
The integral up to $t\sim1/\qtf$ therefore contains the term
\begin{equation}
  \frac{3\ln\rs}{2\pi^3}\int h^2\,\dd z=\frac{1-\ln2}{\pi^2}\ln\rs ,
\end{equation}
which gives the coefficient $A=(1-\ln2)/\pi^2\approx\valA$\,Ha of Gell-Mann and Brueckner. This simple calculation
contains the mechanism of every logarithm below. Screening cuts off the soft region at $q\sim\qtf\propto\rs^{1/2}$, and
the unscreened tail between $\qtf$ and $k_F$ produces $\ln\rs$.

\subsection{Logarithms from the soft region}
\label{sec:regions}

Consider a correction to the polarization of order $e^{2j}$ whose expansion at fixed $z$ contains a term $q^m$.
For one such insertion, the term of order $c^n$ in the expansion of Eq.~\eqref{eq:ring} has radial measure $q^{1+m-2n}\dd q$.
When $m=2n-2$, this gives $\dd q/q$ and hence a term proportional to $\rs^{j+m/2}\ln\rs$, provided its coefficient
survives integration over $z$. An additional factor of $\ln q$ gives a double logarithm~\cite{Imoto2026}.
The $\rs\ln\rs$ term comes from $q^2L_1$ and the long-wavelength limit of $\Pi^{(1)}$.
At order $\rs^2$, the bare-ring contribution involves both $L_2$ and $L_1^2$.
A double logarithm requires a
$q^2\ln q$ term in $\Pi^{(1)}$ or a $\ln q$ term in $\Pi^{(2)}$ at fixed $z$.

The nonlogarithmic coefficients and the single logarithms also need the constants of the screened outer line. With
$G(c)=1/(1+c)-1$ and $c=h/t^2$, the two integrals that enter are Hadamard finite parts,
\begin{equation}
  \mathrm{FP}\!\int_0^\infty t^3G\,\dd t=-\frac{h^2}{2}\ln h,\qquad
  \mathrm{FP}\!\int_0^\infty t\,G\,\dd t=\frac h2\ln h .
\end{equation}
The finite parts remove the large-$t$ tails, $-ht+h^2/t$ in the first integral and $-h/t$ in the second. The
logarithmic tails are removed only for $t>1$. With $\ln\qtf=\tfrac12(\ln\rs+\ln\atf)$, every logarithm of $\rs$ comes
with a term in $\ln\atf$ and a term in $\ln h$, which appear in all coefficients below.

\subsection{Weights of the diagrams}
\label{sec:weights}

Equation~\eqref{eq:ring} gives each closed diagram the weight it has as a ring of proper insertions, once per way of
cutting it into such a ring. We counted the closed-diagram weights directly. The weight of a topology is
$N/(n!\,2^n)$, where $N$ is the number of the $(2n)!$ Wick contractions of $n$ interactions that produce it. A brute-force
enumeration for $n\le5$ gives $1/(2n)$ for the ring of $n$ bubbles, $1/2$ for the exchange diagram with any nonempty
chain of bubbles on one of its lines, and $1$ for the Fock-dressed bubble with any number of plain bubbles. These are
the weights that Eq.~\eqref{eq:ring} assigns. The insertions of Fig.~\ref{fig:diagrams}(b) with Aslamazov--Larkin or
genuine second-order exchange structure, and the iterated first-order insertions, can be cut into a ring in only one
way. They also enter with the weight of Eq.~\eqref{eq:ring}.

The inner class [Fig.~\ref{fig:diagrams}(b)] is different. When bubbles are inserted on both lines of the exchange
diagram, the diagram is a ring with momentum $\bm q$ whose insertion has a screened internal line with momentum $\bm k$.
It is also a ring of the same form with $\bm q$ and $\bm k$ interchanged, and Eq.~\eqref{eq:ring} counts it twice. The
enumeration confirms that all such diagrams have exactly half the ring-formula weight. We therefore use the ordered
insertion, in which the internal momentum is larger than the ring momentum. For the double logarithm the ordering makes
no difference, because the exchange of the two lines maps one region onto the other. For the single logarithm it
defines the finite part $\bin(z)$ of Sec.~\ref{sec:E1}.

\section{The linear coefficient $D$}
\label{sec:D}

\subsection{Three classes of diagrams}

We follow the classification of Carr and Maradudin and of Endo \emph{et al.}~\cite{Carr1964,Endo1999}. Three classes
of diagrams contribute at order $\rs$, giving the estimate
\begin{equation}
  D=D_\mathrm{RPA}+D_{X1}+\text{$E_{32}/\rs$}.
\end{equation}

Bare rings give $D_\mathrm{RPA}$, and rings with one first-order polarization insertion $\Pi^{(1)}$ give $D_{X1}$.
The insertion includes both vertex and self-energy terms. The third class comprises the third-order diagrams whose
three Coulomb lines carry distinct momenta. Their energy per electron $E_{32}$ is proportional to $\rs$
[Eq.~\eqref{eq:E32energy}], so their contribution to $D$ is $E_{32}/\rs$.
For the doubly screened exchange diagram, the symmetrized four-point loop with two soft lines vanishes for a linear
dispersion. The next term is odd in the frequencies and integrates to zero, so this diagram gives no term linear in $\rs$.

The two ring series follow from Eq.~\eqref{eq:ring} by the separation into regions of Sec.~\ref{sec:regions},
\begin{align}
  D_\mathrm{RPA}&=\frac{6\atf}{\pi^3}\int\dd z\Big[-\tfrac12h^2L_1\ln(\atf h)+H(z)\Big],\\
  H(z)&=\int_0^\infty\dd q\,\frac{L^3-h^3-3h^2L_1q^2\,\theta(1-q)}{3q^3},\\
  D_{X1}&=\frac{6\alpha}{\pi^4}\int\dd z\Big[-\tfrac12hS\ln(\atf h)+H_1(z)\Big],\label{eq:DX1}\\
  H_1(z)&=\int_0^\infty\frac{\dd q}{q}\Big[L(q,z)P(q,z)-h(z)S(z)\,\theta(1-q)\Big].
\end{align}
Here $P(q,z)=\pi^3\Pi^{(1)}(q,iqz)/e^2$ is the exact first-order exchange polarization at all momenta. Its
long-wavelength limit is DuBois' function~\cite{DuBois1959}
\begin{equation}
  S(z)=-\frac{(1+3z^2)-(2+3z^2)z\arctan(1/z)}{1+z^2},
  \label{eq:S}
\end{equation}
with $S(0)=-1$, the exchange correction to the compressibility (Sec.~\ref{sec:pol}). Replacing $\ln\atf$ by $\ln\rs$
in the first terms of $D_\mathrm{RPA}$ and $D_{X1}$ gives the ring and exchange parts of $C$,
$\alpha(\pi^2-6)/(24\pi^3)\approx\valCRPA$\,Ha and $A_{12}=\alpha(\pi^2-12\ln2)/(4\pi^3)\approx\valCX$\,Ha. This shows
that the soft regions are treated consistently.

The ring part has a closed form except for one one-dimensional integral. The hard part, $(6\atf/\pi^3)\int H\,\dd z$,
reduces to a single $q$ integral of logarithms and dilogarithms, because $\int L^3\,\dd z$ at fixed $q$ can be done in
closed form. Integrating term by term gives
\begin{multline}
  \frac{\alpha}{2\pi^3}\Big[\frac{943}{120}-\frac{2\pi^2}{9}-5\ln2-\frac{\pi^2}{6}\ln2-\frac{111}{80}\zeta(3)\Big]\\
  \approx\DRhard\Ha .
\end{multline}
The remaining soft part is $(\atf/4\pi^3)\int h^2\ln h\,(1+z^2)^{-2}\,\dd z$. The ring series gives
$D_\mathrm{RPA}=\DRPA$\,Ha, in agreement with a direct fit of the ring energy at small $\rs$ and with the value
$\DRPAendo$\,Ha of Endo \emph{et al.}~\cite{Endo1999}. The exchange ring series needs $P(q,z)$ at all momenta,
including the logarithmic singularity at $q=2k_F$ that a finite frequency rounds off (Appendix~\ref{app:pi1}). It gives
$D_{X1}=\DXone$\,Ha. Table~\ref{tab:D} lists the pieces.

\begin{table}
  \caption{Contributions to the coefficient $D$ of $\rs$, in Ha per electron. The soft parts are the terms with $\ln\atf$ and
  $\ln h$ of Eq.~\eqref{eq:DX1} and its analog for the ring series. The last column gives the corresponding terms of
  Carr and Maradudin~\cite{Carr1964} and the values of Endo \emph{et al.}~\cite{Endo1999}, converted from Ry.}
  \label{tab:D}
  \begin{ruledtabular}
  \begin{tabular}{lrr}
  Contribution & This work & Refs.~\cite{Carr1964,Endo1999} \\
  \hline
  ring, $\ln\atf$ & $\DRlnA$ & \\
  ring, $\ln h$ & $\DRlnh$ & \\
  ring, hard & $\DRhard$ & \\
  $D_\mathrm{RPA}$ & $\DRPA$ & $\DRPAendo$ \\
  \hline
  exchange ring, $\ln\atf$ & $\DXlnA$ & $\cmlnA$ \\
  exchange ring, $\ln h$ & $\DXlnh$ & $\cmlnh$ \\
  exchange ring, hard & $\DXhard$ & $\cmhard$ \\
  $D_{X1}$ & $\DXone$ & $\cmepspp$ \\
  \hline
  $E_{32}/\rs$ & $\Ethree$ & $\Eendo$ \\
  \hline
  $D$ & $\Dnew$ & $\Dold$ \\
  \end{tabular}
  \end{ruledtabular}
\end{table}

\subsection{Third-order diagrams with three different momenta}

Third-order Rayleigh--Schr\"odinger theory with the kinetic energy as unperturbed Hamiltonian gives nine terms that
contribute to $E_{32}$. These are the particle--particle, hole--hole and particle--hole terms, each with a direct and an
exchange part, and a term with the first-order self-energy~\cite{Endo1999}. We write $E_{32}=3\alpha I\rs/(64\pi^7)$ in
Ha, with $I=I_\mathrm{dir}+I_\mathrm{ex}$. The three direct terms are the three time orderings of one Feynman diagram, two triangles connected by three
Coulomb lines with different momenta. In this form the triangle can be done in closed form up to one Feynman-parameter
integral, and the direct terms become a five-dimensional integral that is finite without any regulator. Adaptive Monte
Carlo integration~\cite{Lepage2021} gives $I_\mathrm{dir}=\Idir$. Of the six exchange-type terms, the three
particle--hole terms and the self-energy term reduce to four- and five-dimensional integrals that we evaluate by
deterministic quadrature. The particle--particle and hole--hole exchange terms are done by Monte Carlo. The sum of all six is
$I_\mathrm{ex}=\Iex$ (Appendix~\ref{app:E32}). Together,
\begin{equation}
  E_{32}=\Ethree\,\rs\Ha ,\label{eq:E32energy}
\end{equation}
compared with the value $\Eendo\,\rs$\,Ha of Endo \emph{et al.}, which was obtained with a screened interaction and
given to two figures~\cite{Endo1999}. Excluding only configurations with all three line momenta below a common sharp cutoff changes the direct triangle integral by a term
quadratic in the cutoff, so the extrapolation to zero cutoff is regular. A Yukawa regulator with screening wave number $\kappa$ instead produces
terms of the form $\kappa^2\ln^2\kappa$ from nested small-momentum regions and is not suitable for extrapolation.

\subsection{The coupling-constant weighting of Carr and Maradudin}

The three classes give
\begin{equation}
  D=\Dnew\Ha ,
  \label{eq:D}
\end{equation}
compared with $D\approx\Dold$\,Ha in the literature~\cite{Endo1999}. Table~\ref{tab:D} shows that most of the difference lies in
the exchange ring series. Carr and Maradudin's three terms correspond one to one to ours. Their soft $\ln h$ and hard
terms add up almost to ours. Their $\ln\atf$ term contains the extra constant $-A_{12}/3\approx-\Atwelvethird$\,Ha per electron, where $A_{12}$ is the
exchange part of $C$. This term is not a numerical error.

Carr and Maradudin wrote the exchange ring series as a coupling-constant integral, their Eq.~(19), which they took from
DuBois (his Eq.~(C16), with the opposite overall sign)~\cite{Carr1964,DuBois1959}. Write the $\rs$-dependence of the
exchange ring series as $\epsilon''(\rs)=\rs\phi(\rs)$. Differentiating every Coulomb line
with respect to the coupling parameter $\lambda$ gives the exact identity
\begin{equation}
  \epsilon''(\rs)=\int_0^1\dd\lambda\Big[3\lambda\,\epsilon''(\lambda\rs)+\lambda^3\rs^2\phi'(\lambda\rs)\Big].
\end{equation}
Equation~(19) of Carr and Maradudin keeps only the first term. DuBois obtained the factor of three from two equivalent interaction lines and a contribution from another term in his Eq.~(C2).
For longer bubble chains, our line count shows that the reduction to his Eq.~(C15) omits cuts between adjacent bubbles. A member
of the series with $m$ Coulomb lines is then weighted by $3/m$ instead of 1. In Mellin language the weight is
$M(p)=\int_0^13\lambda^{p+1}\dd\lambda=3/(p+2)$ for a term $\rs^p$. The normalization $M(1)=1$ leaves the lowest
member and the coefficient of $\rs\ln\rs$ unchanged, which is why the check of Carr and Maradudin on the lowest diagram
passes. The slope $M'(1)=-1/3$ acts on the logarithm, $\phi=A_{12}\ln\rs+\dots$, and adds the constant
$-A_{12}/3$. The term $-1/3$ already appears in DuBois' Eq.~(C21), and it passed through Carr and Maradudin into the
value of Endo \emph{et al.}

The Wick counts of Sec.~\ref{sec:weights} settle the question directly. The exchange diagram with two or three bubbles on
one line has weight $1/2$, and the Fock-dressed bubble with two or three further bubbles has weight 1. These are the
ring-formula weights. Equation~(19) of Carr and Maradudin gives these diagrams $3/4$ and $3/5$ of their weight. The
correct $\rs$ coefficient of the exchange ring series is therefore $D_{X1}$ of Eq.~\eqref{eq:DX1}. The remaining
difference of about $\cmNumDiff$\,Ha is based on the values printed by Carr and Maradudin, whose soft $\ln h$
integral is $\cmlnhLow$\,\% low in magnitude. The larger third-order term $E_{32}/\rs$ adds about $\EthreeShift$\,Ha.

\section{The terms of order $r_s^2$}
\label{sec:rs2}

\subsection{Polarization at long wavelength}
\label{sec:pol}

At long wavelength the proper polarization is the response of quasiparticles at the Fermi surface. To first order in
$e^2$ the Landau function is the exchange interaction between equal spins,
$f(\hat{\bm p},\hat{\bm p}')=-2\pi e^2/(1-x)$ with $x=\hat{\bm p}\cdot\hat{\bm p}'$, together with the Fock correction of the
Fermi velocity. We use $f(x)=\sum_l(2l+1)f_l P_l(x)$. The Legendre components $f_l$ diverge because of the forward singularity at $x=1$. Galilean invariance
fixes the velocity through $v^*=1-Nf_1$, with $N=1/2\pi^2$ the density of states per spin. The response therefore depends
only on the differences $f_l-f_1=2\pi e^2(H_l-1)$, which are finite ($H_l$ is the harmonic number). The result is
DuBois' function $S(z)$ of Eq.~\eqref{eq:S}.

\begin{figure}
  \includegraphics[width=\columnwidth]{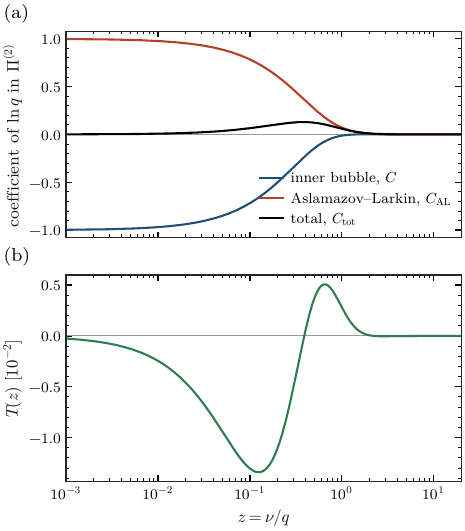}
  \caption{Coefficients of the logarithms of the polarization at fixed $z=\nu/q$. (a) Coefficients of $\ln q$ in the
  second-order polarization, Eq.~\eqref{eq:pi2}, from the inner bubble, $C$, and from the Aslamazov--Larkin diagrams,
  $\CAL$, and their sum $\Ctot$. The two cancel in the static limit, as the compressibility sum rule requires.
  (b) Coefficient $T$ of $q^2\ln q$ in the first-order exchange polarization, Eq.~\eqref{eq:T}.}
  \label{fig:logs}
\end{figure}

The Landau description treats the exchanged Coulomb line as a point in angle. In fact the line
$1/|\bm p-\bm k|^2=1/[(p-k)^2+2pk(1-\cos\theta)]$ between quasiparticles at momenta $\bm p$ and $\bm k$ distinguishes
angles $\theta$ of order $q$, where the radial difference $p-k$, itself of order $q$, is comparable to the angular one.
Separating the angular integral into the region $\theta\gg q$ and the region $\theta\sim q$ gives a $\ln q$ from their
overlap. At order $q^0$ the logarithms of the vertex and of the self-energy cancel, as the Ward identity requires. At
order $q^2$ they do not. We write the first- and second-order polarization at frequency $iqz$ as
\begin{align}
  \frac{\pi^3\Pi^{(1)}}{e^2}&=S(z)+q^2\big[T(z)\ln q+U(z)\big]+O(q^4\ln q),\label{eq:pi1}\\
  \pi^3\Pi^{(2)}&=\frac{2e^4}{\pi}\Big[\Ctot(z)\ln q+B_2(z)\Big]+O(q).\label{eq:pi2}
\end{align}
The coefficient of $q^2\ln q$ in the first-order polarization is
\begin{equation}
  T(z)=-\frac{1}{48}\int_{-1}^{1}\frac{X^4(3X^2-1)}{(X-iz)^4}\,\dd X .
  \label{eq:T}
\end{equation}
Here $X$ is the electron momentum component along $\bm q$ in the limit $q\to0$ (Appendix~\ref{app:pi1}). This form follows from writing the vertex and the Fock term together as one integral, in which the exchange kernel acts
only on differences of the two quasiparticle responses. The integral is elementary (Appendix~\ref{app:pi1}). The
function vanishes in the static limit, $T(0)=0$, in agreement with the static polarization of Engel and Vosko, which has
no $q^2\ln q$ term~\cite{Engel1990}. It decays like $-1/(105z^4)$, as the f-sum rule requires
[Fig.~\ref{fig:logs}(b)]. The logarithm is a three-dimensional effect of the forward exchange. The constant $U(z)$ of
the same order has the static limit $U(0)=1/36$, again the value of Engel and Vosko~\cite{Engel1990}.

At order $e^4$ three terms of the Landau function could produce an infrared logarithm. Two of them do.

The first is the inner bubble. A polarization bubble on the internal line of $\Pi^{(1)}$ gives a Landau function
with a double forward pole, $\delta f\propto\qtf^2/(1-x)^2$. Since $P_l(x)=1-\tfrac12l(l+1)(1-x)+\dots$, the
combination $f_l-f_1$ now diverges logarithmically with an $l$-dependent coefficient, which the velocity
renormalization cannot remove. The divergence is cut off at $\theta\sim q$ and gives $\ln q$ with the coefficient
\begin{align}
  C(z)&=\tfrac12\avg{(1-\mu^2)g'^2}-\avg{g^2}\nonumber\\
  &=\frac{6z(z^2+1)^2\arctan(1/z)-6z^4-10z^2-3}{3(z^2+1)^2}.
\end{align}
Here $g(\mu)=\mu/(\mu-iz)$ is the free response of a quasiparticle at $\mu=\hat{\bm p}\cdot\hat{\bm q}$, and
$\avg{F}=\tfrac12\int_{-1}^1F\,\dd\mu$ is the average over the Fermi surface. The static value $C(0)=-1$ is the
compressibility change produced by a Yukawa-screened exchange.

The second is the Aslamazov--Larkin term. The direct second-order energy, differentiated twice with respect to the
occupations of two different quasiparticles, gives a Landau function that is logarithmically divergent in the window
$q,\qtf\ll k\ll1$ of the internal momentum $k$. Its angular kernel couples only antipodal points of the Fermi surface,
\begin{equation}
  f_\mathrm{AL}(\hat{\bm p},\hat{\bm p}')=-2e^4\ln\frac{1}{k_\mathrm{min}}\,\delta(1+\hat{\bm p}\cdot\hat{\bm p}'),
\end{equation}
so that its Legendre components are proportional to $(-1)^l$. We derived this result for all $l$
(Appendix~\ref{app:pi2}). Only even $l$ survive in $f_l-f_1$, and with $k_\mathrm{min}\to q$,
\begin{align}
  \CAL(z)&=\sum_{l\,\mathrm{even}}(2l+1)G_l^2=\int_0^1\frac{\mu^4\,\dd\mu}{(\mu^2+z^2)^2}\nonumber\\
  &=\frac{2+3z^2-3z(z^2+1)\arctan(1/z)}{2(z^2+1)},
\end{align}
where $G_l=\tfrac12\int_{-1}^1P_lg\,\dd\mu$. Its static value $\CAL(0)=1$ cancels that of the inner bubble. The total
\begin{equation}
  \Ctot=C+\CAL=\frac z2\,\frac{(z^2+1)^2\arctan(1/z)-z^3-\frac53z}{(z^2+1)^2}
\end{equation}
vanishes at $z=0$ [Fig.~\ref{fig:logs}(a)]. This is required by the compressibility sum rule, because the
compressibility follows from the energy, and $A\ln\rs$ in $\epsilon_c$ produces no $\rs^2\ln\rs$ term in the static
polarization.

The third candidate, the genuine second-order exchange, has no logarithm. A soft line can connect two holes, two
particles, or a hole and a particle, and the contributions of these placements cancel. The iterated first-order term
contains only the finite differences $f_l-f_1$ and has no logarithm either. The finite part of Eq.~\eqref{eq:pi2} is the
sum $B_2=\bin+\tfrac\pi2\cAL+B_{2,\mathrm{it}}+B_{2,x2}$ of the inner class, the Aslamazov--Larkin diagrams, the
iterated first-order term and the second-order exchange.

\subsection{The single logarithm}
\label{sec:E1}

Inserted in Eq.~\eqref{eq:ring}, the expansions~\eqref{eq:pi1} and~\eqref{eq:pi2} give the single logarithm
\begin{multline}
  E_1=E_1^\mathrm{ex}+\frac{6\alpha^2}{\pi^5}\!\int h^2T\ln(\atf h)\,\dd z+\frac{12\alpha^2}{\pi^5}\!\int h^2U\,\dd z\\
  -\frac{3\alpha^2}{\pi^5}\!\int h\Ctot\ln(\atf h)\,\dd z-\frac{6\alpha^2}{\pi^5}\!\int hB_2\,\dd z .
  \label{eq:E1master}
\end{multline}
Here $E_1$ denotes the sum of the seven contributions listed in Table~\ref{tab:E1}. Three are known in closed form. The first, $E_1^\mathrm{ex}$,
collects the terms without $\ln q$. These are the Lindhard $q^4$ term of the ring, $\alpha^2(12\pi^2+35)/(2880\pi^4)$,
the product of $S$ with the Lindhard term $L_1$, and two $\Pi^{(1)}$ insertions in one ring. Their sum is
\begin{equation}
  E_1^\mathrm{ex}=\frac{\alpha^2(3012\pi^2-28495)}{2880\pi^4}\approx\Eex\times10^{-3}\Ha .
\end{equation}
The second is the logarithmic part of $\Ctot$ in Eq.~\eqref{eq:E1master}, whose $\ln\atf$ part is
$\alpha^2(1-2\ln2)\ln\atf/(4\pi^4)$ and whose $\ln h$ part is a one-dimensional integral. The third is the iterated
first-order term. The first-order Landau function acts twice, and its contribution to $B_2$ is
$B_{2,\mathrm{it}}=-\Sigma(z)/2$ with
\begin{equation}
  \Sigma(z)=\frac{2+22z^2+15z^4-(6z+27z^3+15z^5)\vartheta-2z^2\vartheta^2}{2(1+z^2)^2},
\end{equation}
where $\vartheta=\arctan(1/z)$ and $\Sigma(0)=1$ is the Hartree--Fock compressibility. The $z$ integral is
elementary and gives
\begin{align}
  E_1^\mathrm{it}&=\frac{\alpha^2[120(1-\ln2)-5\pi^2+12\pi^2\ln2-54\zeta(3)]}{16\pi^4}\nonumber\\
  &\approx\Eit\times10^{-3}\Ha .
\end{align}

\begin{table}
  \caption{Contributions to the coefficient $E_1$ of $\rs^2\ln\rs$, Eq.~\eqref{eq:E1master}, in units of
  $10^{-3}$\,Ha. The uncertainty of the total is discussed in the text.}
  \label{tab:E1}
  \begin{ruledtabular}
  \begin{tabular}{lr}
  Contribution & Value \\
  \hline
  without $\ln q$ & $\Eex$ \\
  \quad (Lindhard $q^4$\ term of $L^4$, $SL_1$, $\Pi^{(1)}\Pi^{(1)}$) & \\
  $q^2\ln q$ and $q^2$ terms of $\Pi^{(1)}$ ($T$, $U$) & $\ETU$ \\
  logarithms of $\Pi^{(2)}$ ($\Ctot$) & $\ECt$ \\
  inner class ($\bin$) & $\Ein$ \\
  Aslamazov--Larkin ($\cAL$) & $\EAL$ \\
  iterated first order ($B_{2,\mathrm{it}}$) & $\Eit$ \\
  second-order exchange ($B_{2,x2}$) & $\Exx$ \\
  \hline
  $E_1$ & $\Eone$ \\
  \end{tabular}
  \end{ruledtabular}
\end{table}

The other four classes need the finite parts $U(z)$, $\bin(z)$, $\cAL(z)$ and $B_{2,x2}(z)$. We computed them by
separating each insertion into a hard region, where the Landau description holds with the exact static functions, and a
soft region at internal momenta of order $q$ (Appendix~\ref{app:pi2}). For $\Pi^{(1)}$ the exact polarization at finite
$q$ gives $U(z)$ directly, to $10^{-7}$. For the Aslamazov--Larkin and inner classes a direct evaluation at finite $q$
with the exact triangles and internal lines, extrapolated to $q\to0$, agrees with the separation into regions to about
$10^{-5}$ for $z\ge0.18$.

\begin{figure*}
  \includegraphics[width=0.85\textwidth]{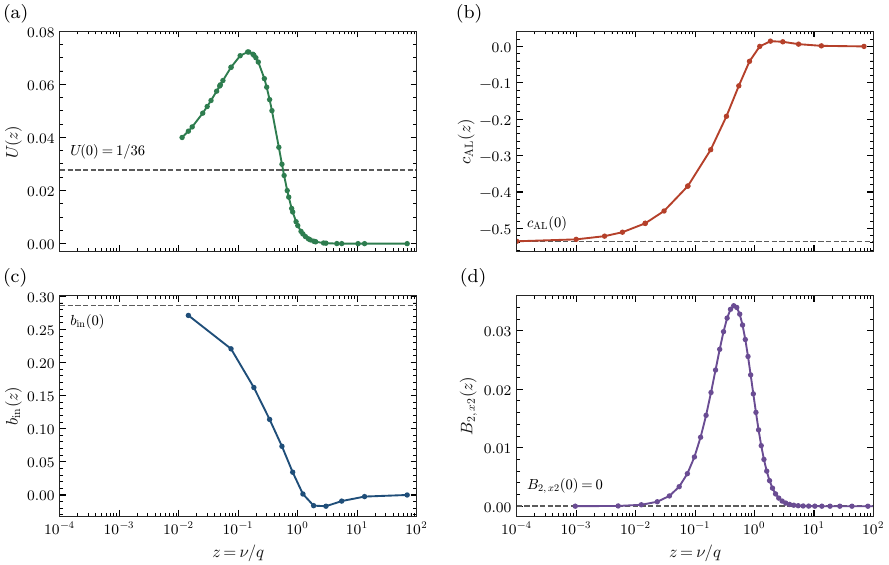}
  \caption{Finite parts of the polarization insertions at fixed $z$, in the normalization of Eqs.~\eqref{eq:pi1}
  and~\eqref{eq:pi2}. (a) Constant $U$ of the $O(q^2)$ term of $\Pi^{(1)}$. (b) Aslamazov--Larkin finite part $\cAL$.
  (c) Finite part $\bin$ of the inner class with the ordered insertion. (d) Second-order exchange term $B_{2,x2}$. The
  dashed lines mark the static limits. The symbols are the computed values, at the nodes of the $z$ quadrature and at
  additional small $z$.}
  \label{fig:finite}
\end{figure*}

Figure~\ref{fig:finite} shows the four functions. All vary on the scale $z\lesssim0.1$, where the density fluctuation is
slow compared with the Fermi velocity. The constant $U$ tends to
$1/36$, the static value of Engel and Vosko. The Aslamazov--Larkin and inner finite parts tend to
$\cAL(0)\approx\cALzero$ and $\bin(0)=\bzero$. The second-order exchange term $B_{2,x2}$ approaches zero quadratically in $z$. Its vanishing static value follows from an exact sum rule, because the second-order exchange energy per electron does not depend on the density, which forces the $l=0$ and
$l=1$ components of its Landau function to be equal. The function $B_{2,x2}$ comes from a Landau function with a
$\ln^2(\pi-\theta)$ singularity at backscattering, which the angular quadrature must resolve.

With these functions, Eq.~\eqref{eq:E1master} gives
\begin{equation}
  E_1=(\EoneLong\pm\EoneMeth)\times10^{-3}\Ha .
  \label{eq:E1}
\end{equation}
The numerical error, from the quadratures and the Monte Carlo integration, is $\EoneNum\times10^{-3}$\,Ha. The
uncertainty we give is larger. It allows for a common error of 1\,\% in $\cAL$ and $\bin$ at the two lowest nodes of Table~\ref{tab:nodes} ($z<0.1$). At these nodes, separate finite-$q$ extrapolations agree with the values obtained by separation into regions.
These two nodes contribute almost half of the Aslamazov--Larkin and inner classes. The largest positive term is the
Aslamazov--Larkin class, and the inner class partly cancels it.

\subsection{The double logarithm}
\label{sec:E2}

The same insertions give the double logarithm,
\begin{equation}
  E_2=\frac{3\alpha^2}{\pi^5}\int h^2T\,\dd z-\frac{3\alpha^2}{2\pi^5}\int h\Ctot\,\dd z ,
  \label{eq:E2master}
\end{equation}
and the terms with $\ln\atf$ in Eq.~\eqref{eq:E1master} add up to $2E_2\ln\atf$. With the functions of
Sec.~\ref{sec:pol} the integrals can be done in closed form,
\begin{equation}
  E_2=\frac{\alpha^2(36\pi^2-208\ln2-221)}{192\pi^4}\approx\Etwo\times10^{-4}\Ha .
  \label{eq:E2}
\end{equation}

\begin{table}
  \caption{Contributions to the coefficient $E_2$ of $\rs^2\ln^2\rs$, Eq.~\eqref{eq:E2master}, in units of
  $10^{-4}$\,Ha, with $\alpha=(4/9\pi)^{1/3}$. The rows are the $q^2\ln q$ term $T$ of $\Pi^{(1)}$, the inner bubble
  and the Aslamazov--Larkin (AL) diagrams.}
  \label{tab:E2}
  \begin{ruledtabular}
  \begin{tabular}{lll}
  Source & Closed form & Value \\
  \hline
  $T$ & $\alpha^2(36\pi^2-160\ln2-245)/(192\pi^4)$ & $\EtwoT$ \\
  inner & $\alpha^2(7-8\ln2)/(8\pi^4)$ & $\EtwoIn$ \\
  AL & $3\alpha^2(\ln2-1)/(4\pi^4)$ & $\EtwoAL$ \\
  \hline
  $E_2$ & $\alpha^2(36\pi^2-208\ln2-221)/(192\pi^4)$ & $\EtwoLong$ \\
  \end{tabular}
  \end{ruledtabular}
\end{table}

Table~\ref{tab:E2} lists the three contributions. The term from $\Pi^{(1)}$ is small, because $36\pi^2-160\ln2-245$
is close to zero. The inner and Aslamazov--Larkin terms are each several times larger than the total and nearly
cancel. This cancellation reflects the vanishing of $\Ctot$ at small $z$, where the weight $h$ of the outer line is
largest.

The coefficient is complete if no other region of the diagrams gives two nested logarithms. A $\ln^2\rs$ needs two
nested logarithmic integrations, $\qtf\ll q\ll k\ll1$. Cutting the softest line leaves a proper insertion with all
internal momenta between $q$ and $1$, and at order $\rs^2$ the only such insertions are those of
Eqs.~\eqref{eq:pi1} and~\eqref{eq:pi2}, which we have computed. The region where two loop momenta are both of order
$\qtf$ has a single scale and gives at most a single logarithm. A double logarithm could still come from a doubly
logarithmic singularity of the symmetrized fermion loops at degenerate configurations. The symmetrized three-point
loop is not bounded. Where its three momenta become collinear and the ratios $\eta$ of frequency to momentum vanish
together, its deviation from the static value $4\pi$ behaves as $\eta^{-1}\varphi(\beta/\eta)$, with $\beta$ the angle
between the momenta, $\varphi(0)$ finite and $\varphi(x)\sim x^{-3}$. This singular set has codimension four in the
two-loop integral, and the singularity is integrable even for the product of two loops in the Aslamazov--Larkin
diagrams. It therefore gives neither $\ln\rs$ nor $\ln^2\rs$. The completeness of Eq.~\eqref{eq:E2} rests on the
assumption that all singularities of the symmetrized loops are integrable in this sense. We tested it on the
three-point loop but did not prove it in general.

\section{Discussion}
\label{sec:discussion}

\begin{figure}
  \includegraphics[width=\columnwidth]{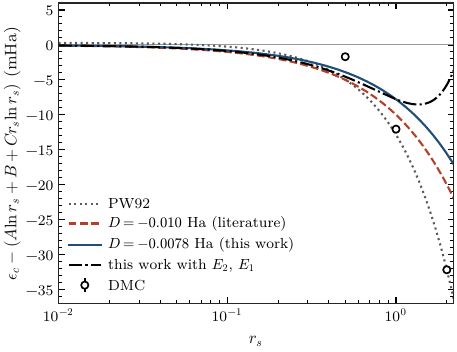}
  \caption{Correlation energy beyond the first three terms of the high-density expansion. The lines show the
  truncation after $D\rs$ with the literature value $D=\Dold$\,Ha~\cite{Endo1999} and with the value of this work, and
  the latter with the two $\rs^2$ logarithms added. The symbols are the DMC energies of Spink
  \emph{et al.}~\cite{Spink2013} and the dotted line is the parametrization PW92~\cite{Perdew1992}, both after
  subtraction of the same three terms.}
  \label{fig:dmc}
\end{figure}

The three-class estimate of $D$ is about \DchangeFrac\ less negative than the literature value. Figure~\ref{fig:dmc} compares the
truncated expansions with the DMC energies of Spink \emph{et al.}~\cite{Spink2013}. At $\rs=0.5$ the two values of $D$
change the energy by only about $\dmcSplit$\,mHa. The DMC energy of Spink \emph{et al.} lies above every truncation shown in the figure, by about
$\dmcAboveD$\,mHa from the expansion through $D$, and also lies above the parametrization PW92~\cite{Perdew1992}. At the same density, Azadi \emph{et al.}~\cite{Azadi2023} find an unpolarized DMC energy 2.5\,mHa per electron below the expansion through our estimate of $D$. The present DMC energies
therefore cannot decide between the two values of $D$. At $\rs\ge1$ the truncated series has not converged, and the DMC
energies follow PW92, which was fitted to quantum Monte Carlo data at these densities.

The figure also shows the behavior of PW92 at high density. As $\rs\to0$ it lies about $\pwHigh$\,mHa above the exact
expansion, because its constant term is slightly too small in magnitude. PW92 therefore underestimates the magnitude of
the correlation energy of the dense gas. Its coefficient of $\rs\ln\rs$ also differs from the exact $C$.

The double logarithm changes the structure of the expansion more than its size. Its term is about a tenth of the
single-logarithm term at $\rs=0.1$ and a fifth at $\rs=0.01$, and it dominates only at extremely high density. Both
terms vanish at $\rs=1$.

\section{Conclusions}

In conclusion, we evaluated contributions to $D$ and to the logarithmic terms of order $\rs^2$.
The three evaluated classes contribute $\Dnew$\,Ha per electron toward $D$. The literature
value $\Dold$\,Ha contains the spurious constant $-A_{12}/3$ from the coupling-constant weighting of the exchange ring
series, which goes back to DuBois. The ring part of $D$ is known in closed form up to one integral.
For the single logarithm, the seven evaluated contributions give $E_1=(\EoneLong\pm\EoneMeth)\times10^{-3}$\,Ha per electron.
Whether these sums exhaust $D$ and $E_1$ remains open. The expansion also contains a double logarithm. Under the integrability assumption discussed in Sec.~\ref{sec:E2}, its coefficient is
$E_2=\alpha^2(36\pi^2-208\ln2-221)/(192\pi^4)\approx\Etwo\times10^{-4}$\,Ha. It arises from three infrared logarithms of
the polarization at long wavelength, namely the forward exchange in the first-order polarization, the screening of its
internal line, and the Aslamazov--Larkin diagrams, whose kernel couples antipodal points of the Fermi surface. The next
step is the nonlogarithmic coefficient of order $\rs^2$, which would
give parametrizations of the correlation energy a firmer basis at high density.

\section*{Acknowledgments}
P.R.P. and M.A.L.M. were supported by a collaboration between the Kavli Foundation, Klaus Tschira Stiftung, and Kevin
Wells, as part of the SuperC collaboration, and by the Simons Foundation through the Collaboration on New Frontiers in
Superconductivity (Grant No.\ SFI-MPS-NFS-00006741-10).

\section*{AI Use Disclosure}
We used Claude Code (Anthropic) and Codex (OpenAI) to help with the derivations, numerical calculations, and drafting of the manuscript.
We take responsibility for the results and conclusions.

\appendix

\section{Lindhard function at fixed $z$}
\label{app:lindhard}

The Lindhard function at imaginary frequency $iqz$ has a regular expansion in $q^2$ at fixed $z$, Eq.~\eqref{eq:lindhard},
with
\begin{equation}
  L_k(z)=-\frac{\mathrm{Re}\big[(2k-iz)(1+iz)^{2k}\big]}{2^{2k}(2k-1)(2k)(2k+1)(1+z^2)^{2k}},\qquad k\ge1 .
\end{equation}
The series converges for $q<2\sqrt{1+z^2}$. The first terms are $L_1=-1/[12(1+z^2)^2]$,
$L_2=(5z^2-1)/[240(1+z^2)^4]$ and $L_3=-(35z^4-42z^2+3)/[6720(1+z^2)^6]$. Inserted in Eq.~\eqref{eq:ring}, $L_1$ gives the
ring part of $C$. The ring part of $E_1$ comes from the $q^4$ term in $L^4$. Both terms in its coefficient, $4h^3L_2+6h^2L_1^2$, contribute to the result
$\alpha^2(12\pi^2+35)/(2880\pi^4)\approx\EoneRPA\times10^{-4}$\,Ha.
A stable numerical evaluation of $L$ switches to this series at small $q$ and to an inverse series at large $|q/2+iz|$.

\section{First-order polarization at finite momentum}
\label{app:pi1}

At finite $q$ we write the vertex part of $\Pi^{(1)}$ as a two-dimensional integral over the momentum components along
$\bm q$, with the transverse integrals done in closed form, and the Fock part as a one-dimensional integral. Composite
Gauss--Legendre rules graded toward the Fermi surface give $P(q,z)$ to about $10^{-13}$ for $q\ge0.01$. The static
limit reproduces the series of Engel and Vosko~\cite{Engel1990}, $(P+1)/q^2=1/36+O(q^2)$. The hard part of $D_{X1}$ uses
$P$ on a $q$ grid clustered at $q=2$ up to $q=10$, where the integrand has fallen below $10^{-8}$. The step of the
subtraction term at $q=1$ is integrated analytically. Two $z$ rules with 36 and 42 nodes agree to $10^{-9}$\,Ha.

At long wavelength we combine the vertex and the Fock term into one integral,
$\pi^3\Pi^{(1)}/e^2=\tfrac1{16}\iint w(X)w(Y)k_q(X,Y)\,\dd X\,\dd Y$ with $w=1/(X-iz)^2$, $X=p_\parallel+q/2$, $Y=k_\parallel+q/2$, and a kernel
$k_q$ that does not depend on $z$ and acts only on differences. Expanding $k_q$ in $q$ gives $S(z)$ at order $q^0$ and,
from the region where $X-Y$ is of order $q$, the coefficient $T(z)$ of Eq.~\eqref{eq:T}. Its closed form is
\begin{multline}
  T(z)=-\frac{5z}{2(z^2+1)^3}\Big[\big(z^2+\tfrac1{15}\big)(z^2+1)^3\arctan\frac1z\\
  -z^7-\tfrac{41}{15}z^5-\tfrac{107}{45}z^3-\tfrac35z\Big].
\end{multline}
The constant $U(z)$ is the sum of an elementary term, a term with dilogarithms and a two-dimensional integral. It agrees
with fits of the exact $P(q,z)$ at small $q$ to $10^{-7}$.

\section{Second-order insertions}
\label{app:pi2}

\emph{Aslamazov--Larkin kernel.} With $\omega=ks$ and $k\ll1$, the derivative of the free polarization with respect to
the occupation of a quasiparticle at $\hat{\bm p}$ becomes $(is-\hat{\bm p}\cdot\hat{\bm k})^{-2}$. Here $\mu=\hat{\bm p}\cdot\hat{\bm k}$ and $\mu'=\hat{\bm p}'\cdot\hat{\bm k}$. The kernel
$J_0=\int\dd\Omega_k\int\dd s\,(is-\mu)^{-2}(is-\mu')^{-2}$ can be written with the representation
$(i\omega-a)^{-2}=-\int_0^\infty\tau e^{-|\omega|\tau}e^{\mp i\tau a}\dd\tau$. After the frequency and momentum
integrals, the result is supported where the two Fermi velocities are antiparallel, and a local analysis near that point
gives $J_0=4\pi^2\delta(1+\hat{\bm p}\cdot\hat{\bm p}')$. With $x=\hat{\bm p}\cdot\hat{\bm p}'$, the Legendre expansion is $J_0(x)=4\pi\sum_l(2l+1)a_l P_l(x)$, with
$a_l=\int\dd s\,d_l(s)^2=(-1)^l\pi/2$ for all $l$. Here $d_l(s)=-Q_l'(is)$.

\emph{Finite parts.} The Aslamazov--Larkin finite part is the sum of a hard term, a Legendre sum of the static Landau
function with $z$-independent coefficients, and a soft term written as Fermi-surface averages of the small-momentum
triangles. The inner class uses the ordered insertion with the dynamic internal line. Its hard part is the Landau theory
with the exact static Lindhard function plus the dynamic remainder of the $GW$ velocity, and its soft part is the
$O(q^0)$ term of the insertion at internal momenta of order $q$. The static limits satisfy the identity
$L^\mathrm{in}(0)+\tfrac\pi2L^\mathrm{AL}(0)=-(1-\ln2)$ between the hard parts, which follows from the compressibility sum rule
with a sharp infrared cutoff on the Coulomb lines. The finite-$q$ evaluations of the Aslamazov--Larkin and inner
diagrams use the exact triangles, a momentum grid that resolves the jump of the occupations at $K=2$, and extrapolation
in $q$ at fixed $z$.

The second-order exchange term $B_{2,x2}$ is the Landau projection of the Landau function obtained from the
second-order exchange energy, $f_{x2}=\partial^2E_{2x}/\partial n_a\partial n_b$, with the exact Galilean term. A soft
region of the iterated and second-order exchange classes could add a constant at $O(q^0)$. It vanishes, because each
exchange line enters through differences of quasiparticle quantities and the closed loops with three or more density
vertices cancel for a linear dispersion. Where both lines are soft, only the sum of these classes and the Fock-type
self-energies cancels, which is what $E_1$ needs.

Table~\ref{tab:nodes} gives the finite parts at the 12 nodes of the $z$ quadrature. The nodes are $z=\tan\phi$ with
Gauss--Legendre nodes in $\phi\in(0,\pi/2)$, and $w$ is the weight for the full real axis. The values of $E_1$ in
Table~\ref{tab:E1} use converged rules with 24 to 48 nodes. On the 12-node rule each class changes by less than
$10^{-6}$\,Ha.

\begin{table*}
  \caption{Finite parts of the polarization insertions at the nodes of the 12-point $z$ rule, in the normalization of
  Eqs.~\eqref{eq:pi1} and~\eqref{eq:pi2}. The numerical uncertainties are below $10^{-7}$ for $U$, $2\times10^{-5}$ for
  $\cAL$, $4\times10^{-5}$ for $\bin$ and $10^{-7}$ for $B_{2,x2}$. The iterated term $B_{2,\mathrm{it}}=-\Sigma/2$ is
  exact.}
  \label{tab:nodes}
  \begin{ruledtabular}
  \begin{tabular}{rrrrrrr}
\hline\hline
$z$ & $w$ & $U$ & $c_\mathrm{AL}$ & $b_\mathrm{in}$ & $B_{2,\mathrm{it}}$ & $B_{2,x2}$ \\
\hline
$0.0145$ & $0.0741$ & $0.042341$ & $-0.48639$ & $0.27121$ & $-0.46686$ & $0.000324$ \\
$0.0754$ & $0.1689$ & $0.066389$ & $-0.38390$ & $0.22074$ & $-0.34750$ & $0.005733$ \\
$0.1827$ & $0.2598$ & $0.070955$ & $-0.28394$ & $0.16223$ & $-0.20252$ & $0.019187$ \\
$0.3360$ & $0.3552$ & $0.054224$ & $-0.19207$ & $0.11404$ & $-0.09163$ & $0.031787$ \\
$0.5418$ & $0.4744$ & $0.029897$ & $-0.10837$ & $0.07341$ & $-0.03218$ & $0.033295$ \\
$0.8204$ & $0.6547$ & $0.011940$ & $-0.04089$ & $0.03429$ & $-0.00883$ & $0.023855$ \\
$1.2190$ & $0.9729$ & $0.003664$ & $0.00012$ & $0.00133$ & $-0.00185$ & $0.011833$ \\
$1.8458$ & $1.6163$ & $0.000932$ & $0.01455$ & $-0.01628$ & $-0.00027$ & $0.003932$ \\
$2.9765$ & $3.1465$ & $0.000177$ & $0.01279$ & $-0.01712$ & $-2.13\times10^{-5}$ & $0.000801$ \\
$5.4731$ & $7.7836$ & $0.000018$ & $0.00632$ & $-0.00938$ & $-6.55\times10^{-7}$ & $0.000082$ \\
$13.2540$ & $29.6768$ & $5.69\times10^{-7}$ & $0.00165$ & $-0.00255$ & $-3.47\times10^{-9}$ & $2.52\times10^{-6}$ \\
$69.0452$ & $353.3406$ & $7.83\times10^{-10}$ & $9.78\times10^{-5}$ & $-0.00015$ & $-1.76\times10^{-13}$ & $3.46\times10^{-9}$ \\
\hline\hline
\end{tabular}

  \end{ruledtabular}
\end{table*}

\section{Third-order terms}
\label{app:E32}

The nine terms of $E_{32}$ give $I=\sum c_t\int\dd^{12}p\,\theta/(L_1^2L_2^2L_3^2D_1D_2)$. Here $\dd^{12}p$ denotes integration over four independent three-dimensional momenta. Each of the three Coulomb lines contributes a factor $1/L_i^2$, where $L_i$ is the magnitude of its momentum transfer. The occupation factors are collected in $\theta$, and $D_1$ and $D_2$ are the kinetic-energy denominators of the two intermediate states~\cite{Endo1999}. Table~\ref{tab:E32} gives the signed weight $c_t$ of each term $t$. In the direct triangle the three oriented line momenta sum to zero. In the
Feynman form of the direct terms, $I_\mathrm{dir}=F/(12\pi^2)$, the triangle is reduced to a closed-form ball integral
and one Feynman parameter. Near a vertex where both momentum and frequency are small, the residue terms of the triangle
cancel to high order, and a Monte Carlo estimate of this region has heavy tails. We exclude the region where both are
below $\eta$ at one vertex and correct for it with its $O(\eta^2\ln\eta)$ dependence. The uncertainty of
$I_\mathrm{dir}$ covers this correction and the extrapolation of a cutoff series to zero cutoff. Table~\ref{tab:E32} gives the terms. The second-order exchange
energy, computed with the same programs, reproduces the closed form of Ref.~\cite{Onsager1966} to $10^{-9}$.

\begin{table*}
  \caption{Terms of the third-order integral $I$, with $E_{32}=3\alpha I\rs/(64\pi^7)$ in Ha, and the method used for
  each term.}
  \label{tab:E32}
  \begin{ruledtabular}
  \begin{tabular}{llrl}
\hline\hline
Term & Coefficient & Contribution to $I$ & Method \\
\hline
direct (pp, hh, ph) & $+2,+2,-4$ & $970.47(10)$ & Monte Carlo, Feynman form \\
pp exchange & $-1$ & $-975.449(41)$ & Monte Carlo \\
hh exchange & $-1$ & $-504.526(6)$ & Monte Carlo \\
ph exchange, two equivalent terms & $+2$ each & $1037.9956\times2$ & quadrature \\
ph exchange, third term & $+2$ & $283.2598$ & quadrature \\
self-energy & $+1$ & $-874.4149$ & quadrature \\
\hline
exchange sum & & $4.86(4)$ & \\
$I$ & & $975.33(11)$ & \\
\hline\hline
\end{tabular}

  \end{ruledtabular}
\end{table*}

\bibliography{references}

\end{document}